\documentclass[12pt]{iopart}
\usepackage[modulo]{lineno}
\usepackage{hyperref}
\usepackage{subfloat}
\usepackage{subcaption}
\usepackage{amssymb}
\usepackage{epstopdf}
\usepackage{graphicx}
\usepackage[usenames, dvipsnames]{xcolor}

\colorlet{red}{black}
\colorlet{blue}{black}

\begin{document}

\title{The intermediate scattering function for quasi-elastic scattering in the presence of memory friction}

\author{Peter S.M. Townsend$^1$, David J. Ward$^1$}
\address{$^1$ Cavendish Laboratory, J.J. Thomson Avenue, Cambridge CB3 0HE, UK}
\ead{psmt2@cam.ac.uk}

\begin{abstract}
We derive an analytical expression for the intermediate scattering function of \textcolor{red}{a particle on a flat surface} obeying the Generalised Langevin Equation, with exponential memory friction. Numerical simulations based on an extended phase space method confirm the analytical results. The simulated trajectories provide qualitative insight into the effect that introducing a finite memory timescale has on the analytical line shapes. The relative amplitude of the long-time exponential tail of the line shape is suppressed, but its decay rate is unchanged, reflecting the fact that the cutoff frequency of the exponential kernel affects short-time correlations but not the diffusion coefficient which is defined in terms of a long-time limit. \textcolor{blue}{The exponential sensitivity of the relative amplitudes to the decay time of the chosen memory kernel is a very strong indicator for the prospect of inferring a friction kernel and the physical insights from experimentally measured intermediate scattering functions.}
\end{abstract}


\submitto{\textcolor{blue}{Journal of Physics Communications}}
\maketitle


\section{Introduction}

Noise and dissipation play a crucial role in the nature and the rate of dynamical processes at surfaces. For example, the interplay of noise, dissipation and the average interaction potential controls surface diffusion. It is crucial to understand these fundamental dynamical behaviours which are necessary for the self-assembly of clusters \cite{Cui2011ChemComm}, nanotubes \cite{Hofmann2005PRL} and nano-structured arrays \cite{Pawin2006Science} if one wants to employ a bottom up approach in building nanoscale devices. In equilibrium, noise and dissipation are linked by a fluctuation-dissipation relation \cite{Kubo1966RepProgPhys}. The simplest phenomenological model for classical, dissipative surface diffusion in equilibrium is the Langevin equation \cite{MiretArtes2005JPCM}, in which the coupling of translational motion to the thermal bath is represented by white noise (fluctuation) and a friction parameter (dissipation). \textcolor{red}{Langevin descriptions of ion motion based on the coupling to electronic excitations are by no means limited to the study of adsorbate dynamics at surfaces. For example the concept finds application in high energy density physics \cite{Dai2010PRL} where it has recently been used to model nuclear fusion experiments within a multiscale framework \cite{Stanton2018PRX}. However, the present work will focus entirely on dynamical properties of adsorbates on surfaces in thermal equilibrium.} The rate and detailed nature of thermally driven diffusion is well known to depend strongly on the magnitude of the friction \cite{Hanggi1990RevModPhys}. In the case of dissipation by the creation and annihilation of electron-hole pairs, the friction can be estimated from first principles \cite{Juaristi2008PRL} and comparison with experiment then allows the contribution of other dissipation channels to be estimated \cite{Rittmeyer2016PRL}.

In real systems the noise spectrum is generally not white, but is suppressed at high frequencies and can be described using a cutoff frequency $\omega_{c}$, above which the power density is zero or falls rapidly to zero. For example in microcanonical molecular dynamics simulations of an adsorbate on a harmonic solid substrate, the numerically derived cutoff frequency is finite \cite{Gershinsky1996SurfSci}. Any coloured noise spectrum defines a Generalized Langevin Equation (GLE). According to the GLE, the position $x(t)$ of a particle of mass $m$ in one dimension, in thermal equilibrium at temperature $T$ with a linearly responding bath with memory obeys
\begin{equation}
\label{eqn:GLE}
m\ddot{x}(t)=-\nabla V(x)-\int^{t} m\gamma(t-t')\dot{x}(t')dt'\, + F(t)\,\textrm{,}
\end{equation}
where $V(x)$ is a static potential energy landscape, $\gamma(t)$ is a causal function (i.e. with a factor of $\theta(t)$, the Heavisde step function) known as the friction kernel, and $F(t)$ is a self-correlated random force. The fluctuation-dissipation relation, and prescriptions for the lower limit of the friction integral, will be specified in Section \ref{sec:analytic}. The details of the friction kernel and the statistics of the associated noise are highly dependent on the specific system studied. However, there are many reasons to consider the effect of a simple analytical noise spectrum. For one thing, representing the friction kernel compactly with a small number of parameters means that the parameter space can be feasibly explored. In addition, specific choices of the friction kernel can lead to greatly simplified analytical and numerical methods. 

A kernel decaying exponentially in time $\gamma(t)=\gamma\omega_{c}\theta(t)e^{-\omega_{c}t}$ is often the simplest choice, because its Laplace transform has only a single, simple pole. The interpretation of the exponential decay rate $\omega_{c}$ as a cutoff frequency is justified later on via the connection of $\gamma(t)$ with either the spectral density of a system/reservoir coupling (\ref{eqn:lorentzianCoupling}) or equivalently the power spectrum of the random force $F(t)$ (\ref{eqn:fourierFDT}). \textcolor{blue}{In some model systems $\gamma(t)$ is not well described by an exponential decay, for example the motion of a tracer particle harmonically bound in a one-dimensional harmonic chain. In that case, $\gamma(t)$ has an algebraically decaying envelope \cite{Rubin1963PhysRev,Maekawa1980PLA,Mokshin2005NJP} associated with a hard frequency cutoff in the Fourier transform $\tilde{\gamma}(\omega)$. However, the results of molecular dynamics simulations do not conform to the simple one-dimensional harmonic chain result \cite{Gershinsky1996SurfSci} and there are potential contributions to the friction kernel from additional effects such as inter-adsorbate interactions within an interacting single adsorbate framework \cite{MartinezCasado2007JPCM}, that would likely smooth out a hard frequency cutoff. \textcolor{red}{In liquid-phase many-body simulations \cite{Harp1970PRA,Gertner1989JCP} the effective friction kernel is typically not a perfect exponential, but the non-exponential behaviour is not due to a hard cutoff frequency. Therefore for a mobile adsorbate on a real surface there is no strong reason to prefer an algebraic-envelope model over an exponential model \textit{a priori}. The essential deviation from Langevin behaviour is then captured by the finite width of the kernel in time, not by the exact shape of the decay.} The exponential model gives the simplest possible functional representation of a kernel that does not decay to zero so quickly that it can be treated as a $\delta$-function.} The Langevin equation corresponds to the limiting case $\omega_{c}\rightarrow \infty$. In activated jump diffusion, $\omega_{c}$ has an impact on the jump rates, both in the moderate-to-high friction regime \cite{Grote1980JCP} and the energy diffusion limited regime \cite{Hershkovitz1999SurfSci}. However in practice we note that to some extent a trade-off between the values of $\gamma$ and $\omega_{c}$ is possible such that knowledge about jump distributions is not especially diagnostic of $\gamma$ or $\omega_{c}$ individually. Therefore we turn our discussion away from relatively coarse features of surface diffusion such as jump rates, towards equilibrium correlation functions and their experimental measurement.

The effects of a finite $\omega_{c}$ could in principle be seen in experimental measurements that quantify surface diffusion on all timescales, from the timescale of inter-well jumps down to the timescale of ballistic motion and intra-well diffusion. The ballistic-to-Brownian transition has been studied experimentally for mesoscopic particles in a liquid environment \cite{Huang2011NatPhys, Pusey2011Science}, via the velocity autocorrelation function (VACF). Comparable measurements at the atomic scale can be provided by the helium-3 surface spin echo (HeSE) technique \cite{Jardine2009ProgSurfSci}. HeSE broadly provides the experimental motivation behind the present work. To date the Langevin equation has been used extensively to model HeSE measurements of adsorbate dynamics. The frictional coupling $\gamma$ can be quantified by optimising the simulation to reproduce inelastic \cite{Lechner2015PCCP} or quasi-elastic \cite{Hedgeland2009NatPhy} features in the experimental line shapes. The general theory behind 
the HeSE measurements has been set out and reviewed previously \cite{Jardine2009ProgSurfSci}. HeSE line shapes can typically be interpreted in terms of the intermediate scattering function (ISF) for adsorbate motion, denoted $I(\Delta K,t)$. We define $I(\Delta K,t)$ here for one adsorbate as the classical function
\textcolor{blue}{\begin{equation}
\label{eqn:ISF}
I(\Delta K,t)=\langle e^{i\Delta K\cdot{x}(t)} e^{i\Delta K\cdot{x}(0)}\rangle_{\beta}\,\textrm{,}
\end{equation} }
where $\Delta \mathbf{K}$ is the quasielastic momentum transfer, usually varied in the experiment; $\mathbf{x}(t)$ is the classical adsorbate position, $x(0)$ is an initial condition, and $\langle \cdots \rangle_{\beta}$ denotes thermal equilibrium averaging over initial conditions, at an inverse temperature $\beta=(k_{B}T)^{-1}$. For further details on the precise connection between $I(\Delta K,t)$ and HeSE line shapes, we recommend a comprehensive review article \cite{Jardine2009ProgSurfSci}. Here we will refer to the ISF and ``the line shape" interchangeably. 

Analytical line shapes have previously been derived for a range of dynamical models. Notably, the classical ISF is known for a Langevin particle on a flat surface or in a harmonic well \cite{Vega2004JPCM}. Approximate analytical results have been found for the quantum mechanical case too \cite{Vega2004JCP,MartinezCasado2010ChemPhys}, but from here on we confine the discussion entirely to classical dynamics. For Langevin dynamics in a periodic potential when the timescales of intra-well and jump motion are well separated, the relationship between the long-time behaviour of the ISF and the jump dynamics of the system is well known. For example, in the case of a single adsorption site per unit cell, the ISF at long times is an exponentially decaying function of time, and the decay rate has a sinusoidal dependence on $\Delta K$ with the periodicity of the reciprocal lattice \cite{Chudley1961PPSL}. There is no generally applicable analytical result for the ISF in periodic potentials for all $t$, although in some 
regimes an approximate form can be constructed out of the building blocks of analytical vibrational and hopping line shapes \cite{MartinezCasado2007JCP}. To date, simple exact analytical results for the ISF including memory friction have not been published. \textcolor{red}{In our study we will derive such a result. Some readers will recognise a close connection to other types of dynamical correlation function, which are already known for diffusion subject to memory friction \cite{Grabert1988PhysRep}. However, the explicit computation of the ISF, a detailed discussion of the sensitivity of the ISF to memory effects, are both presented here for the first time.}

To make the connection to experimental surface diffusion data in the future, it will necessary to consider a potential energy landscape $V(x)$ with at least a weak corrugation, for which exact analytical line shapes are unlikely to ever be found, and numerical simulation methods are required. A number of methods have been described for simulating the GLE with more or less general friction kernels \cite{Berkowitz1983JCP,Wan1998MolPhys,Bao2004JStatPhys,Stella2014PRB}, and we do not attempt a comprehensive review here, but focus on one method that is convenient for the present application. The extended phase space method of Ceriotti was introduced originally as a method to enhance canonical sampling efficiency in a context where the molecular dynamics are fictitious \cite{Ceriotti2011JCP}, but nevertheless correctly simulates dynamical properties. We will give a succinct account in Section \ref{sec:numerical} of how the method incorporates memory friction. Simulations of exponential memory friction have been performed previously to examine the effect of memory on escape rate \cite{Ianconescu2015JCP}. In contrast, in 
our simulations we are concerned primarily with the ISF at short correlation times, comparison between simulation and analytical theory, and the interpretation of the line shape in terms of the realisations of the GLE generated by the simulation method. \textcolor{red}{In the present study we report only simulations of a particle on a flat surface. However,  the generalisation of the simulation method to include a potential of mean force is very straightforward, so the method is perfectly suitable for the study of memory effects in experimental systems.}

\textcolor{red}{Overall, the aim of our work is to provide a conceptual and practical reference that clearly demonstrates the sensitivity of the ISF to memory friction, which will facilitate the incorporation of memory friction as standard in the simulation and analysis of experimental surface systems. Our use of an exponential kernel is therefore largely motivated by simplicity, allowing us to set out the central ideas concisely and describe the sensitivity to memory effects in a mathematically precise way. We do not claim that the combination of different microscopic dissipation mechanisms leads to a perfect exponential kernel for any specific physical system. However, the simulation method we discuss can be used with modest generalisations to explore memory effects in experimental systems, while the analytical solution for the exponential kernel provides both a useful benchmarking reference and also a conceptual reference for the qualitative effect of memory friction on simulated or experimental correlation functions. Therefore, we expect our study to be a useful building block for researchers studying the influence of memory friction in surface dynamics.}

\section{Analytical results}
\label{sec:analytic}

\subsection{Ensemble average formulation}
\label{sec:ensembleAveraging}

Here we consider the GLE arising from a well-known type of linear system-bath Hamiltonian,
\begin{equation}
\label{eqn:CalLegHam}
H=\frac{p^{2}}{2m}+V(x)+\sum_{\alpha} [\frac{p_{\alpha}^{2}}{2m_{\alpha}}+\frac{1}{2}m_{\alpha}\omega_{\alpha}^{2}(x_{\alpha}-\frac{c_{\alpha}x}{m_{\alpha}\omega_{\alpha}^{2}})^{2}] \, \textrm{,}
\end{equation}
which describes a particle of mass $m$ linearly coupled to a large set of harmonic bath modes. $\alpha$ indexes the modes, which have masses $m_{\alpha}$ and natural oscillation frequencies $\omega_{\alpha}$. $c_{\alpha}$ are the linear coupling coefficients. When considered as a classical Hamiltonian, a textbook derivation \cite{Weiss2012QDS} of the equations of motion reveals that the particle of co-ordinate $q$ obeys the GLE
\begin{equation}
\label{eqn:CalLegGLE}
m\ddot{x}(t)=-V'(x)-\int_{0}^{t} m\gamma(t-t')\dot{x}(t')dt'\, + F(t) \, \textrm{,}
\end{equation}
where the friction kernel is given by
\begin{equation}
\label{eqn:CalLegKernel}
\gamma(t)=\theta(t)\sum_{\alpha}\frac{c_{\alpha}^{2}}{m_{\alpha}\omega_{\alpha}^{2}}\cos(\omega_{\alpha}t)
\end{equation}
and the fluctuating force is given by
\begin{equation}
\label{eqn:CalLegForce}
F(t)=m\sum_{\alpha}[c_{\alpha}(x_{\alpha}(0)-\frac{c_{\alpha}x(0)}{m_{\alpha}\omega_{\alpha}^{2}})\cos(\omega_{\alpha}t)+\frac{p_{\alpha}(0)}{m_{\alpha}\omega_{\alpha}}\sin(\omega_{\alpha}t)]\,\textrm{,}
\end{equation}
which is random because the initial conditions $\{x_{\alpha}(0),p_{\alpha}(0)\}$ are drawn from an initial probability distribution. When the initial distribution represents that of thermal equilibrium, the random force is found to be Gaussian, and has the fluctuation-dissipation property
\textcolor{blue}{\begin{equation}
\langle F(t)F(0) \rangle_{\beta} = m k_{B}T\gamma(t) \, \textrm{,}
\end{equation} }
where the angle brackets represent the canonical average over all possible initial states of the bath. Correlation functions such as \textcolor{blue}{$\langle F(t)F(0) \rangle_{\beta}$} are invariant under a shift of the reference time $t=0$. For the purposes of calculating correlation functions, we will keep $t=0$ as a reference time, and calculate averages based on forward evolution of the system from its initial thermal state. Before beginning the derivation of the ISF, we note that within the present ensemble-average view of the GLE, if we assume a constant density of bath modes in $\omega_{\alpha}$, then the required couplings $c_{\alpha}$ can be specified in a straightforward way to construct an exponential kernel, namely
\begin{equation}
\label{eqn:lorentzianCoupling}
c_{\alpha}^{2}\propto \frac{\omega_{\alpha}^{2}}{\omega_{\alpha}^{2}+\omega_{c}^{2}}\,\textrm{,}
\end{equation}
as can be verified by writing (\ref{eqn:CalLegKernel}) in integral form. In other words, the global Hamiltonian can be parametrised in terms of the cutoff frequency, lending a secondary interpretation to $\omega_{c}$.

We now derive $I(\Delta K,t)$ for an adsorbate obeying the GLE in a one dimensional harmonic well of frequency $\Omega$, for which the GLE reads
\begin{equation}
\label{eqn:harmonicGLE}
m\ddot{x}(t)=-\Omega x(t)-\int_{0}^{t} m\gamma(t-t')\dot{x}(t')dt'\, + F(t) \textrm{.}
\end{equation}
Following \cite{Vega2004JPCM} we write the ISF as a cumulant expansion,
\begin{equation}
I(\Delta K,t)=\exp\Big(-\Delta K^{2}\int_{0}^{t}(t-t')\Psi(t')dt'\Big),
\end{equation}
where \textcolor{blue}{$\Psi(t)=\langle v(t)v(0)\rangle_{\beta}$} is the velocity autocorrelation function. Throughout the present subsection we always interpret \textcolor{blue}{$\langle \rangle_{\beta}$ as an ensemble average at temperature $T$ where $\beta=1/(k_{B}T)$.} The cumulant expansion is exact for flat and harmonic potentials because the probability density for the particle position $x(t)$ is Gaussian at all times.

Within the canonical system-plus-reservoir framework, it is straightforward to prove the absence of certain correlations
\textcolor{blue} {\begin{equation}
\langle F(t)v(0)\rangle_{\beta}=\langle x(t)v(0)\rangle_{\beta}=0 \textrm{.}
\end{equation} }
Because such correlations vanish, Laplace-transforming (\ref{eqn:harmonicGLE}) yields the simple result
\begin{equation}
\hat{\Psi}(s)=\frac{k_{B}T/m}{s+\hat{\gamma}(s)+\Omega^{2}/s}\textrm{,}
\end{equation}
which can be inverse transformed using the Bromwich integral \cite{Riley2006MathMeth}, as long as we can find the roots of the denominator. Now we specialise the GLE (\ref{eqn:GLE}) to the case where the friction kernel $\gamma(t)$ is an exponentially decaying function
\begin{equation}
\label{eqn:expKernel}
\gamma(t)=\theta(t)\gamma\omega_{c}e^{-\omega_{c}t}\,\textrm{,}
\end{equation}
with Laplace transform
\begin{equation}
\hat{\gamma}(s)=\frac{\gamma\omega_{c}}{s+\omega_{c}}\,\textrm{,}
\end{equation}
such that the Laplace transform of the VACF becomes
\begin{equation}
\label{eqn:laplaceVACFHarmonic}
\hat{\Psi}(s)=\frac{s(s+\omega_{c})k_{B}T/m}{s^{3}+\omega_{c}s^{2}+(\gamma\omega_{c}+\Omega^{2})s+\Omega^{2}\omega_{c}}\,\textrm{.}
\end{equation}
The exact inversion of \textcolor{blue}{$\hat{\Psi}(s)$} as specified here would be a natural starting point for an analytical investigation of the dissipation-induced broadening of inelastic line shape features, in the presence of memory friction. Here we focus entirely on quasielastic features on the flat surface $\Omega=0$, in which case all the poles of the denominator have negative real part and the Bromwich integral \cite{Riley2006MathMeth} can be applied straightforwardly with an integration contour consisting of the imaginary axis $\mathcal{IA}$ and a semicircle to the left of $\mathcal{IA}$. The classical ISF is symmetric in $t$ for all $\Delta K$, and all results in the present work are given for $t\geq 0$. When $\Omega=0$ the denominator of (\ref{eqn:laplaceVACFHarmonic}) has two roots, $s_{1}$ and $s_{2}$, which are the solutions of
\begin{equation}
\label{eqn:laplaceRoots}
s^{2}+\omega_{c}s+\gamma\omega_{c}=0 \,\textrm{.}
\end{equation}
The Bromwich formula yields
\begin{equation}
\label{eqn:vacfResult}
\Psi(t)=\frac{k_{B}T}{m} \Big[p_{1}e^{\textcolor{blue}{s_{1}t}}+p_{2}e^{\textcolor{blue}{s_{2}t}}\Big] \, \textrm{,}
\end{equation}
where
\begin{equation}
\label{eqn:pdefs}
p_{1}=\frac{(s_{1}+\omega_{c})}{s_{1}-s_{2}} \, \textrm{ ; } \, p_{2}=\frac{(s_{2}+\omega_{c})}{s_{2}-s_{1}} \,\textrm{.}
\end{equation}
The inversion could alternatively be carried out using the ansatz of a biexponential VACF and matching coefficients after the forward Laplace transform.  The form (\ref{eqn:vacfResult}) is a known result for the VACF \cite{Berne1966JCP}. Integrating (\ref{eqn:vacfResult}) according to the prescription in the cumulant expansion method, we arrive at the ISF in a product form
\begin{equation}
\label{eqn:ISFProductForm}
I(\Delta K,t)=\prod_{k=1}^{2}\exp\Big(-\Delta K^{2}\frac{k_{B}T}{m} \frac{p_{k}}{\textcolor{blue}{s_{k}^{2}}}\Big[e^{s_{k}t}-s_{k}t-1\Big]\Big)\,\textrm{.}
\end{equation}
Each factor in the product has the same functional form as the Langevin flat surface result \textcolor{red}{$I_{L}(\Delta K,t)$} \cite{MiretArtes2005JPCM}, 
\begin{equation}
\label{eqn:ISFLangevinResult}
\textcolor{red}{I_{L}}(\Delta K,t)=\exp\Big(-\Delta K^{2}\frac{k_{B}T}{m\gamma^{2}} \Big[e^{-\gamma t}+\gamma t-1\Big]\Big)\,\textrm{.}
\end{equation}
The Langevin result can be recovered by setting $\omega_{c}\rightarrow\infty$.

Inserting explicit expressions for the roots of (\ref{eqn:laplaceRoots}) yields for the ISF the closed form $I(\Delta K,t)=\exp[-\Delta K^{2}X(t)]$ with
\begin{equation}
\label{eqn:msd}
X(t)=\frac{k_{B}T}{m\gamma^{2}}\Big(\frac{\gamma}{\omega_{c}}-1+\gamma t+\frac{e^{-\omega_{c}t/2}}{\omega_{c}}[C\cosh(\omega't)+S\sinh(\omega't)]\Big)
\end{equation}
where the coefficients $C$ and $S$ are given by
\begin{equation}
C=\omega_{c}-\gamma ;
\end{equation}
\begin{equation}
\textcolor{blue}{S}=\frac{\sqrt{\omega_{c}}(\omega_{c}-3\gamma)}{\sqrt{\omega_{c}-4\gamma}}
\end{equation}
and the frequency scale $\omega'$ mixes the two basic frequency scales of the problem $\omega_{c}$ and $\gamma$ in the following way 
\begin{equation}
\omega'=\frac{1}{2}\sqrt{\omega_{c}^{2}-4\gamma\omega_{c}} \,\textrm{.}
\end{equation}
$\omega'$ becomes complex unless the cutoff frequency $\omega_{c}$ is at least four times faster than the basic strength of the friction $\gamma$. The classical ISF remains real at all times.

\begin{figure}
\centering
\def\svgwidth{0.75\columnwidth}
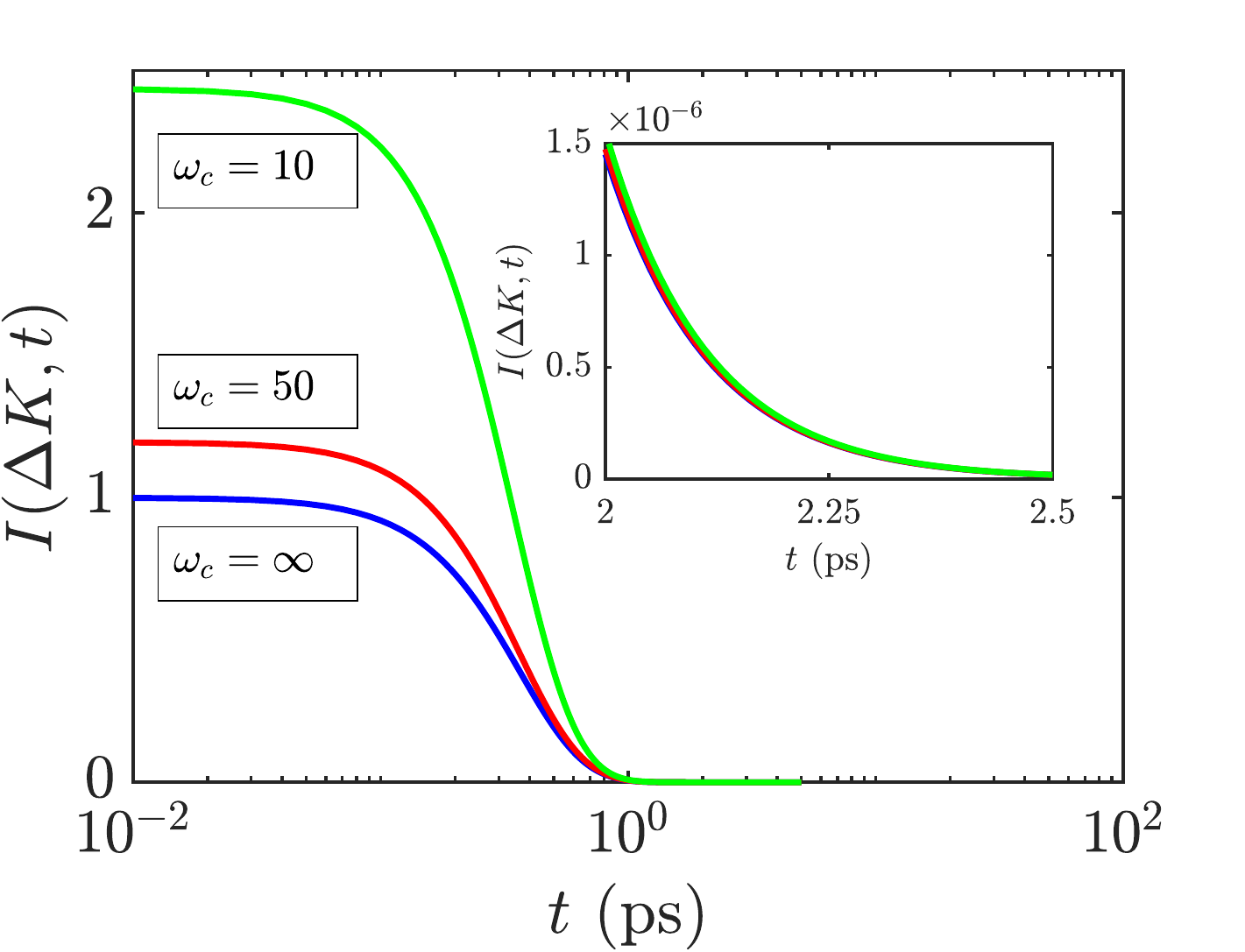
\caption{A series of analytical ISFs at $\Delta K=1.0\,${\AA}$^{-1}$ given by (\ref{eqn:ISFProductForm}), \textcolor{blue}{associated with the GLE with the friction kernel \ref{eqn:expKernel}, where the overall strength of friction is $\gamma=2.0\,$ps$^{-1}$ but different values of the cutoff frequency $\omega_{c}$ are explored.} The ISFs have been normalised to each other at long times, where they all coincide as shown in the inset. At increasingly small values of $\omega_{c}$, the relative amplitude of the fast decay at shorter times increases in accordance with the factor in (\ref{eqn:initialDropSize}). \textcolor{red}{The curve labelled $\omega_{c}=\infty$ is the Langevin result $I_{L}(\Delta K,t)$, given by (\ref{eqn:ISFLangevinResult}).}}
\label{fig:analyticFlatISFs}
\end{figure}

We now compare GLE line shapes for increasing values of the memory time. The results are plotted in Figure \ref{fig:analyticFlatISFs}. The most readily apparent feature is an enhanced initial decay relative to the long-time exponential form. The suppression of the long time tail in the presence of memory is due to the presence of a faster-decaying factor in the general product expression (\ref{eqn:ISFProductForm}). However, the decay rate of the slowest-decaying factor at long times is unchanged by the memory. To be more concrete about the suppression of the long time tail, we note that the long time limit of the Langevin result is
\textcolor{blue}{\begin{equation}
\label{eqn:langevinLongTimeLimit}
I(\Delta K,t\rightarrow \infty)=\exp\Big(-\Delta K^{2}\frac{k_{B}T}{m\gamma^{2}}\Big)\exp\Big(-\Delta K^{2}\frac{k_{B}T}{m\gamma}t\Big) \,\textrm{,}
\end{equation}}
but in the present case with the inclusion of memory friction, \textcolor{blue}{taking the long time limit of \ref{eqn:ISFLangevinResult} shows that} the expression is modified to
\textcolor{blue}{\begin{equation}
\label{eqn:GLELongTimeLimit}
I(\Delta K,t\rightarrow \infty)=\exp\Big(-\Delta K^{2}\Big[\frac{k_{B}T}{m\gamma^{2}}-\frac{k_{B}T}{m\gamma\omega_{c}}\Big]\Big)\exp\Big(-\Delta K^{2}\frac{k_{B}T}{m\gamma}t\Big) \,\textrm{.}
\end{equation} }
\textcolor{blue}{Comparing \ref{eqn:langevinLongTimeLimit} and \ref{eqn:GLELongTimeLimit},} we conclude that the GLE ISF in the long time limit is smaller than the corresponding LE ISF by the constant factor
\begin{equation}
\label{eqn:initialDropSize}
\exp\Big(\Delta K^{2}\frac{k_{B}T}{m\omega_{c}\gamma}\Big) \textrm{\textcolor{blue}{,}}
\end{equation}
\textcolor{blue}{which shows that the relative amplitude of the long time exponential tail of the ISF is exponentially sensitive to $1/\omega_{c}$. The strong functional dependence on $\omega_{c}$ is a key outcome of the present work, and suggests strong prospects for observing memory effects in experimental data under suitable conditions.}

The effect \textit{of finite $\omega_{c}$ on decay amplitudes} is illustrated in Figure \ref{fig:analyticFlatISFs}, which shows the GLE and LE results normalized to each other at long times, for a range of memory frequencies $\omega_{c}$ \textcolor{blue}{ while $\gamma$ is held constant at $2.0\,$ps$^{-1}$}. A notable consequence of the equality of the long-time decay rate for different $\omega_{c}$ is that the diffusion coefficient $D$ is independent of $\omega_{c}$ too, which we demonstrate in one dimension. \textcolor{blue}{At long times the mean square displacement (MSD) is dominated by a linear term, and the diffusion coefficient in one dimension is defined as half the linear coefficient of that term. Owing to the connection between the MSD and the VACF, the MSD is twice the function $X(t)$ given above. Comparing equations \ref{eqn:ISFProductForm} and \ref{eqn:msd}, we find that the linear cofficient in $X(t)$ at long times is
\begin{equation}
D=\frac{k_{B}T}{m}\Big|\Big[\frac{p_{1}}{s_{1}}+\frac{p_{2}}{s_{2}}\Big]\Big|\,\textrm{.}
\end{equation}
Subsituting the relations \ref{eqn:pdefs} and the fact that $s_{1}s_{2}=\gamma\omega_{c}$ courtesy of \ref{eqn:laplaceRoots}, we find that $\omega_{c}$ completely cancels, leaving:
\begin{equation}
D=\frac{k_{B}T}{m\gamma}\,\textrm{.}
\end{equation} }

\textcolor{blue}{Therefore the well-known Stokes-Einstein relation for Brownian motion, which holds for the Langevin equation with friction $\gamma$ \cite{MiretArtes2005JPCM}, also applies for the GLE \ref{eqn:CalLegGLE} with the kernel \ref{eqn:expKernel} for an arbitrary cutoff frequency $\omega_{c}$. The result might at first appear counter-intuitive, but can be understood by noting that the time integral of the friction kernel \ref{eqn:expKernel} is independent of $\omega_{c}$. Consequently, by the fluctuation-dissipation theorem (anticipating the relation \ref{eqn:fourierFDT}), the power density of the random force at $\omega=0$ must also be independent of $\omega_{c}$. Therefore it is entirely reasonable to expect at least some properties of the long-time behaviour to be independent of $\omega_{c}$. The fact that $D$ is independent of $\omega_{c}$ is completely consistent with the ISF at long times having the same decay rate but different amplitude, since adding a constant to $X(t)$ does not change the diffusion coefficient but does rescale the ISF by a constant factor. The ISF at long times therefore contains a footprint of the short time behaviour.}

\textcolor{blue}{As further clarification on the scope of the result, we stress that the diffusion coefficient calculated here applies strictly to the model system in which the potential energy landscape $V(\mathbf{x})$ is perfectly flat. In a periodic potential, additional frequency scales such as the frustrated translation frequencies in local wells, and imaginary oscillation frequencies at saddle points in $V(\mathbf{x})$, will interplay with $\omega_{c}$ leading to a variation of $D$ with $\omega_{c}$.}

When $\omega_{c}<4\gamma$, oscillatory effects are seen due to the complex $\omega'$, even in the absence of a confining potential ($V(x)=0$). The oscillations are a known phenomenon from GLE simulations \cite{Caratti1998ChemPhysLett}. The oscillations are prominent in the VACF, as will be seen in Section \ref{sec:numerical}. However, the effect is subtle in the ISF because the integration and exponentiation steps in the cumulant expansion both smooth out the oscillations.

\subsection{Time average formulation}

We now turn to consider a GLE in which the memory properties, and the correlations of the fluctuating force, are expressed in the Fourier domain via a causal friction kernel and a fluctuation-dissipation theorem in terms of the power spectrum of the random force. We express  correlation functions as time averages, and construct them via the Wiener-Khinchin theorem \cite{Riley2006MathMeth}. \textcolor{blue}{To conceptually distinguish the time averages here from the ensemble averages in Section \ref{sec:ensembleAveraging}, we will write correlation functions in the present subsection without the $\beta$ subscript.}

\textcolor{blue}{Because of the ergodicity and equilibrium properties of the GLE \cite{Ottobre2011Nonlin}, it will not be at all surprising to find that the results of the previous section can be re-stated in the language of Fourier transforms and time averages, rather than Laplace transforms and ensemble averages. Laplace transforms are the easier route to the analytical results, but the numerical method considered later is more conveniently justified in the language of Fourier transforms. Therefore it will be of value to explicitly demonstrate that the analytical results can be derived from a Fourier-space statement of the GLE.} 

We consider the GLE
\begin{equation}
\label{eqn:fourierGLE}
m\ddot{x}(t)=-m\int_{-\infty}^{t} dt' \gamma(t-t')\dot{x}(t') + F(t)
\end{equation}
where now the time integration extends to $-\infty$ and the random force has the power spectrum
\begin{equation}
\label{eqn:fourierFDT}
|\tilde{F}(\omega)|^{2}=2mk_{B}T\tilde{\gamma}'(\omega)
\end{equation}
where $\tilde{\gamma}'(\omega)$ is the real part of the friction kernel. We will also assume that the random force is Gaussian, for consistency with the Hamiltonian model in (\ref{sec:ensembleAveraging}). To find the velocity correlation $\langle v(t)v(0)\rangle$ by time averaging, we construct $|\tilde{v}(\omega)|^{2}$ and make use of the power spectrum of the fluctuating force. Using the convention $\tilde{g}(\omega)=\int dt e^{i\omega t}g(t)$ for the forward Fourier transform, the power spectrum of the velocity becomes:
\begin{equation}
m^{2}|\tilde{v}(\omega)|^{2}=\frac{2mk_{B}T\tilde{\gamma}'(\omega)}{|\tilde{\gamma}(\omega)-i\omega|^{2}} \textrm{.}
\end{equation}
After introducing the exponential kernel, we arrive at
\begin{equation}
m^{2}|\tilde{v}(\omega)|^{2}=\frac{2mk_{B}T\gamma\omega_{c}^{2}}{\omega^{4}+\omega^{2}(\omega_{c}^{2}-2\gamma\omega_{c})+\gamma^{2}\omega_{c}^2} \textrm{.}
\end{equation}
There are four poles in the complex $\omega$ plane, of which two contribute residues in any specific Fourier inversion. We now perform the Fourier inversion assuming $t>0$. Denoting the roots of
\begin{equation}
\label{eqn:fourierRoots}
\omega^{4}+\omega^{2}(\omega_{c}^{2}-2\gamma\omega_{c})+\gamma^{2}\omega_{c}^2=0
\end{equation}
as $\omega_{k}$ and labelling them such that the first two roots ($\omega_{1}$,$\omega_{2}$) have negative imaginary part, the Fourier inversion
\begin{equation}
\langle v(t)v(0)\rangle=\frac{1}{2\pi}\int d\omega e^{-i\omega t}|v(\omega)|^{2}
\end{equation}
can be expressed in terms of those poles, via the residue theorem, as
\begin{equation}
\langle v(t)v(0)\rangle=-2i\frac{k_{B}T\gamma\omega_{c}}{m(\omega_{1}-\omega_{2})} \Big[\frac{e^{-i\omega_{1}t}}{(\omega_{1}-\omega_{3})(\omega_{1}-\omega_{4})}-\frac{e^{-i\omega_{2}t}}{(\omega_{2}-\omega_{3})(\omega_{2}-\omega_{4})}\Big].
\end{equation}
It is easily shown that the solutions of (\ref{eqn:fourierRoots}) satisfy $\omega^{2}=-s^{2}$ where $s$ solves (\ref{eqn:laplaceRoots}). Because the solutions $s_{k}$ derived in the Laplace case have negative real part, we can choose to write
\begin{equation}
\omega_{1}=is_{1};\,\, \omega_{2}=is_{2};\,\, \omega_{3}=-is_{1};\,\, \omega_{4}=-is_{2}\,\textrm{.}
\end{equation}

Substituting the roots $\omega_{k}$ into the Fourier inversion formula, one finds that the correlation function $\langle v(t)v(0)\rangle$ (and therefore $I(\Delta K,t)$) calculated by the time averaging formalism is identical to the one calculated by ensemble averaging. In other words, the device of extending the friction integral to the infinite past, combined with the fluctuation-dissipation relation as given in (\ref{eqn:fourierFDT}) represents exactly the same dynamical equilibrium properties as those represented by Hamiltonian evolution of a system-with-reservoir in a thermal state at a reference time.

In the present section we have derived analytical line shapes within two different formulations of the GLE. The physical interpretation in each case is rather different, in particular the way in which initial conditions of the Brownian particle are handled. We next look for further physical insight into the way the ISF is modified by memory friction, by inspection of numerical trajectories.

\section{Numerical simulation}
\label{sec:numerical}

In the current section we simulate the GLE using the extended phase space prescription of Ceriotti \cite{Ceriotti2011JCP}. \textcolor{red}{Before beginning the mathematical and numerical detail, it is worth briefly placing the GLE in the broader context of isothermal molecular dynamics simulations, and different thermostats such as the Nos{\'e}-Hoover (NH) method \cite{Nose1984JCP,Hoover1985PRL} and velocity rescaling (VR) methods \cite{Bussi2007JCP}. The appropriateness of the thermostat depends strongly on the physical context. For example, one of the features of the cited VR method is that dynamical properties of the simulation are only very weakly dependent on the thermostat parameter. Such a thermostat is therefore highly suitable for many-body simulations in which the interesting dynamics arise from explicit interactions. By contrast, LE or GLE thermostats do affect the dynamical properties of many-body simulations \cite{Basconi2013JCTC}. In the present context, where all interactions of a single simulated particle are represented implicitly via a random force and friction, the strong dependence of dynamical correlations on the thermostat parameters (the friction kernel) is not a limitation but is an essential feature underpinning the use of the LE or GLE to study surface dynamics. Further, the physical connection between the friction kernel and excitations of the environment, as described by (\ref{eqn:CalLegHam}-\ref{eqn:CalLegForce}), implies that we can learn about the dissipative coupling of adsorbates to surface excitations by studying the surface dynamics using the GLE. Within single-particle simulations, comparable information could not be obtained if the system was studied with thermostats that approximately preserve the non-dissipative dynamical information.}

The \textcolor{red}{method we employ} was presented in a matrix notation; here we write out our relevant special case explicitly and show that it reproduces the GLE, (\ref{eqn:fourierGLE}). Suppose our physical variable $x(t)$ is coupled to the motion of a fictitious velocity variable $p(t)$ which undergoes a Langevin process. To simplify the derivation we neglect the potential $V(x)$, which is simply carried through when present. The appropriate choice of coupled equations for an exponential friction kernel and for satisfying the fluctuation-dissipation relation are:
\begin{equation}
\label{eqn:ceriotti1}
m\ddot{x}(t)=\sqrt{\gamma\omega_{c}}mp(t) \, \textrm{;} \,\textrm{,}
\end{equation}
and
\begin{equation}
\label{eqn:ceriotti2}
 m\dot{p}(t)=-\sqrt{\gamma\omega_{c}}m\dot{x}(t)-\omega_{c}mp(t)+\sqrt{2m k_{B}T\omega_{c}}R(t)\,\textrm{,}
\end{equation}
where $R(t)$ is a normalised white noise source. To prove that the simulation prescription achieves the GLE (\ref{eqn:fourierGLE}), we transform the coupled equations to the Fourier domain and eliminate the fictitious variable, and obtain:
\begin{equation}
m\dot{v}(t)=-m\int_{-\infty}^{t}\gamma(t-t')v(t')dt' + \omega_{c}\sqrt{2mk_{B}T\gamma} \, \mathcal{F}^{-1}\Big[\frac{R(\omega)}{\omega_{c}-i\omega}\Big]
\end{equation}
where the power spectrum of the final term has the correct frequency dependence to satisfy the fluctuation-dissipation relation.

\begin{figure}
\centering
\def\svgwidth{0.75\columnwidth}
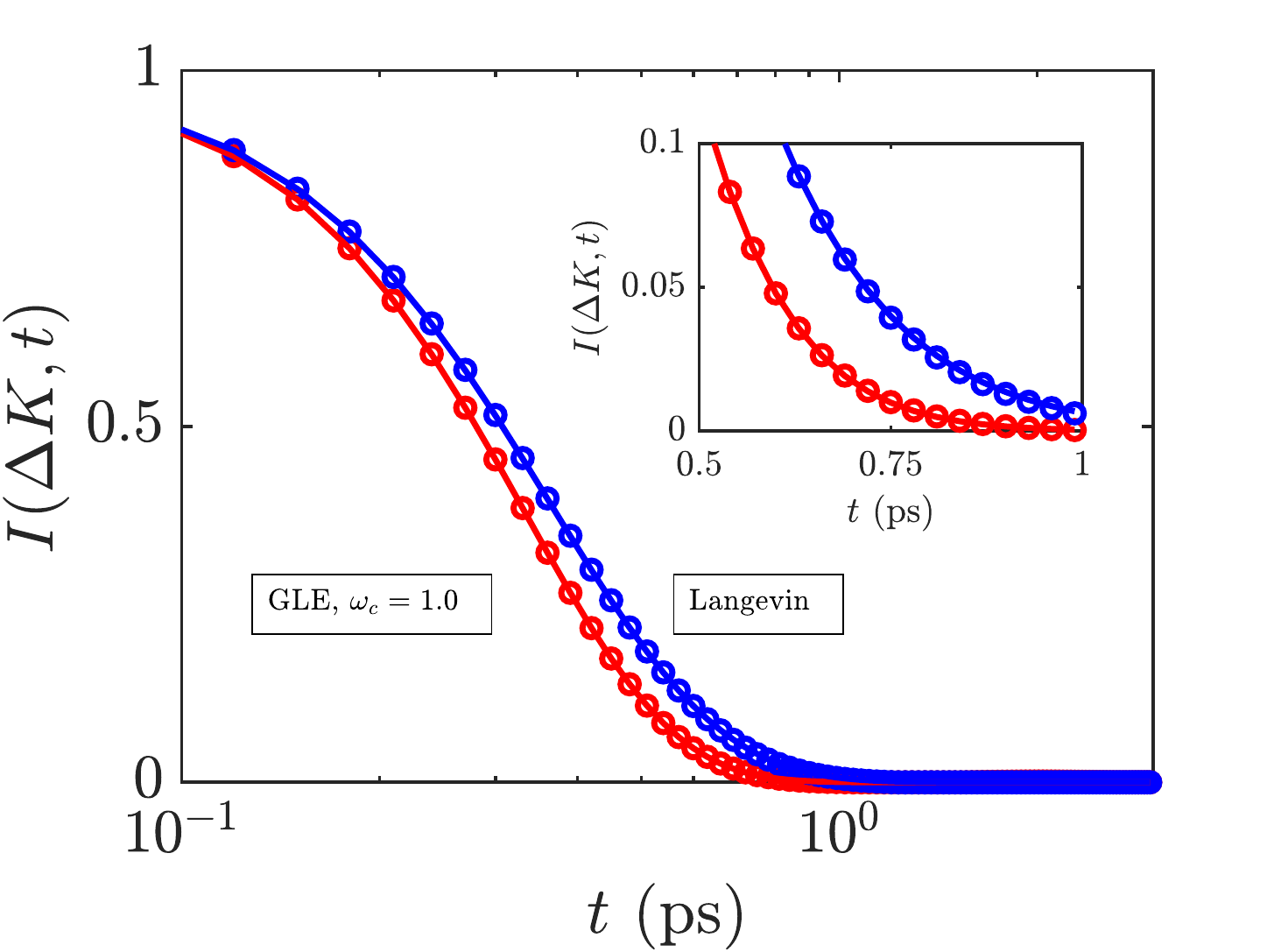
\caption{Simulated ISF (red points) on the flat surface, at $\Delta K=1.0\,${\AA}$^{-1}$. The key simulation parameters are $\gamma=2.0\,$ps$^{-1}$, $\omega_{c}=1.0\,$ps$^{-1}$, such that $\omega_{c}<\gamma$, namely we are well into the regime where memory is important. The analytical result is overlaid (solid red curves) and shows good agreement with the simulation results. The inset demonstrates the agreement of the simulated and analytical results in the regime where the ISF decays exponentially to zero. Additional simulation parameters are specified in the main text. \textcolor{red}{The analogous results for a Langevin simulation at the same value of $\gamma$ are shown in blue. The emergence of a large ratio between the results of the two simulations can be seen clearly in the inset.}}
\label{fig:benchmarkingSimulation}
\end{figure}

To benchmark the correctness of the simulation we simulate the ISF on a flat surface. The coupled equations (\ref{eqn:ceriotti1}) and (\ref{eqn:ceriotti2}) were solved using a velocity Verlet integrator. The resulting simulated line shape at $\Delta K=1.0\,${\AA}$^{-1}$ is shown in Figure \ref{fig:benchmarkingSimulation}, with the analytical result overlaid, showing good agreement. We emphasise that the ISF computed from the simulated trajectories was not constructed via the cumulant expansion, but directly from the definition in (\ref{eqn:ISF}). The simulations were performed for 100 particles with a timestep of $1\,$fs, for a total simulation time of $1\,$ns. The relatively long simulation time means we expect the boundary effects of the finite time domain to be negligible, but additionally we have used a padding scheme to ensure we calculate the linear, not the cyclic correlation, when applying the discrete Wiener-Khinchin theorem. Similarly, the initial conditions were generated with a short period of pre-thermalisation, but since the simulation is much longer than the thermalisation timescale, the exact nature of the pre-thermalisation is not important.

\begin{figure}
\centering
\def\svgwidth{0.75\columnwidth}
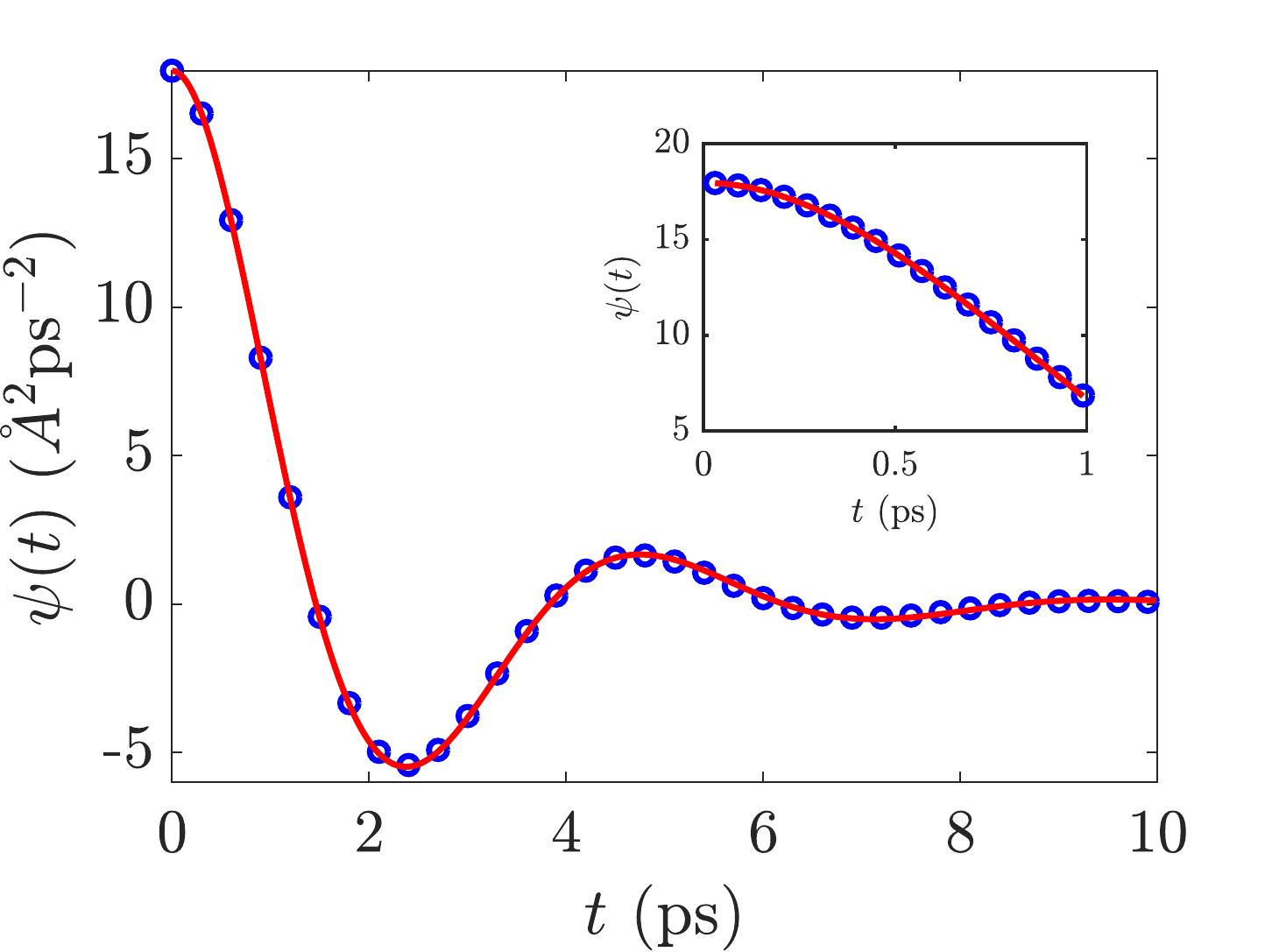
\caption{The velocity autocorrelation function (VACF) $\psi(t)$ on the flat surface, with $\gamma=2.0\,$ps$^{-1}$, $\omega_{c}=1.0\,$ps$^{-1}$. The analytical result in (\ref{eqn:vacfResult}) is shown as a solid red line. The blue points are derived from the same simulated trajectories that gave the simulated ISF in Figure \ref{fig:benchmarkingSimulation}. The inset shows the short-time decay of the VACF in more detail. The limit $\psi(0)$ is simply the mean square velocity at the simulation temperature $T$. Unlike the ISF, $\psi(t)$ exhibits clear oscillations.}
\label{fig:simulatedFourierVACF}
\end{figure}

The simulation parameters associated with the presented line shapes were $\gamma=2.0\,$ps$^{-1}$, $\omega_{c}=1.0\,$ps, $T=150\,$K, $m=7\,$amu. The parameters are typical of those for a light adsorbate moving in a flat landscape, and by simulation with $\omega_{c}<\gamma$ we are very far away from the Markovian limit, such that the VACF displays significant oscillations. The analytical and simulated VACF agree, and are plotted together in Figure \ref{fig:simulatedFourierVACF}.

\begin{figure}
\centering
\def\svgwidth{0.75\columnwidth}
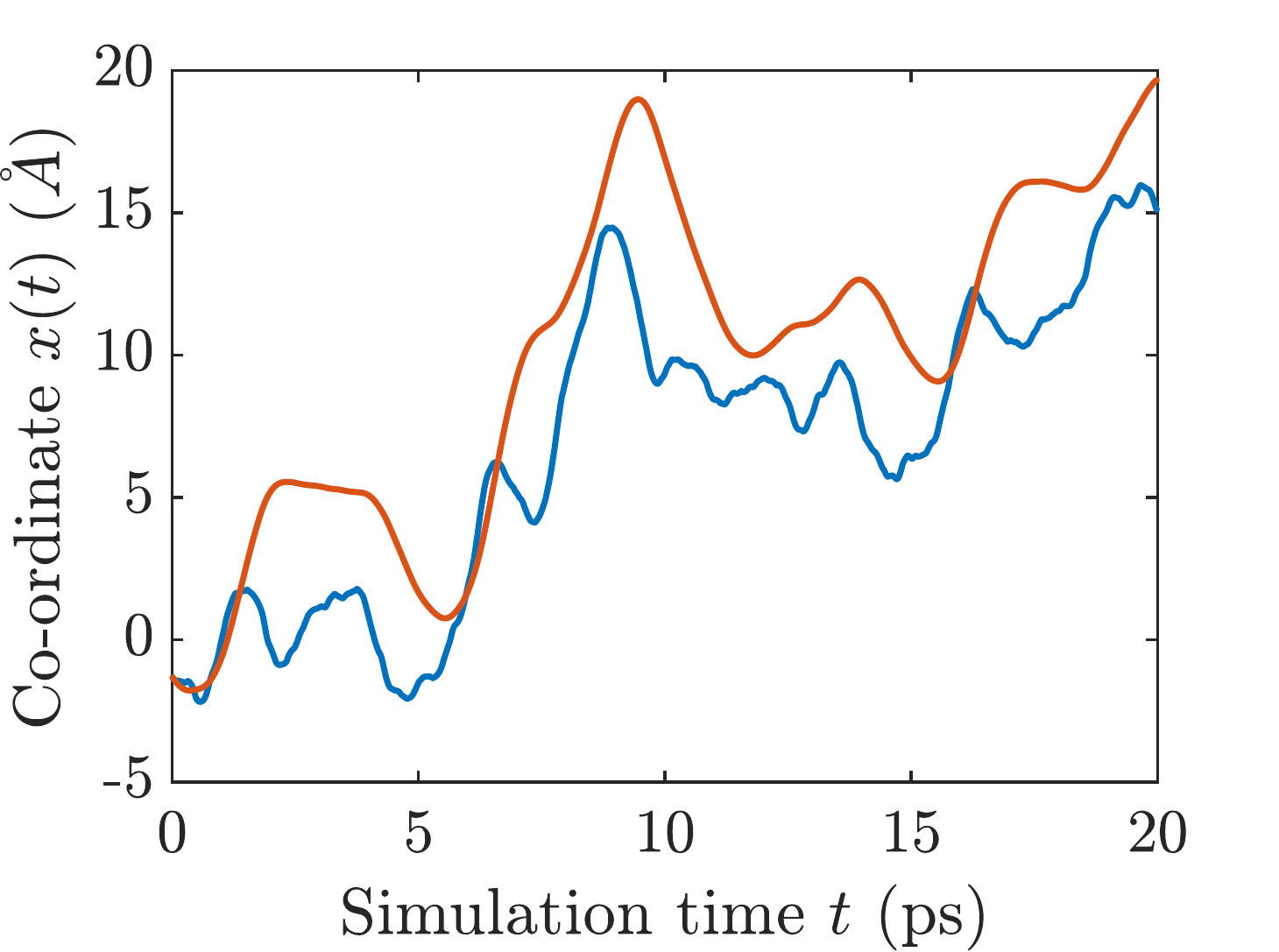
\caption{A $20\,$ps extract of two one-dimensional trajectories on a flat surface. The friction strength is $\gamma=2.0\,$ps$^{-1}$, with memory times of $1.0\,$ps (red) and $0.02\,$ps (blue). The case $\omega_{c}>>\gamma$ (blue) is chosen to be close to the Markovian limit, in which the GLE becomes the Langevin equation. As expected on the basis of the Fourier-space statement of the GLE given in (\ref{eqn:fourierGLE}), the trajectory simulated with a long memory time (low cutoff frequency) is a filtered, or smoothed, version of the Langevin trajectory. The significance of the smoothing in terms of the ISF is discussed in the main text.}
\label{fig:compareTrajectories}
\end{figure}

One of the key results in the present work is that the amplitude of the long-time exponential tail of the ISF, which represents a continuous diffusion process, is reduced by the presence of a finite cutoff frequency in the friction kernel $\gamma(t)$. The corresponding diffusion rate is unchanged, as shown in Section \ref{sec:analytic}. In other words, by the time the diffusive limit of the correlation function is attained, the kinematic scattering amplitude $\exp(i\Delta K\cdot x)$ of our GLE particle is more decorrelated than the same quantity for the corresponding Langevin particle with $\omega_{c}\rightarrow\infty$. The result can be understood qualitatively by inspection of simulated GLE trajectories that are equivalent in all respects other than the memory time. Figure \ref{fig:compareTrajectories} shows two one-dimensional trajectories $x(t)$ simulated using the same initial conditions and the same white noise realisation, differing only in the way the extended phase space formulation filters the 
noise via $\omega_{c}$. The white noise acts directly on the auxiliary variable $p(t)$ of (\ref{eqn:ceriotti2}), but the effect on the physical co-ordinate $x(t)$ is different in each case because of the way $x$ and $p$ are coupled, which depends on $\omega_{c}$. The trajectories are shown over the course of a $20\,$ps time window. One trajectory is simulated in the Markovian limit using a large cutoff frequency $\omega_{c}=500\,$ps$^{-1}$. The other trajectory has $\omega_{c}=1.0\,$ps$^{-1}$ such that memory has a substantial effect on $I(\Delta K,t)$ and $\psi(t)$ as seen in Figures \ref{fig:benchmarkingSimulation} and \ref{fig:simulatedFourierVACF}. The two trajectories approximately track each other over the $20\,$ps timescale plotted, which is expected since the diffusion coefficient is unchanged by the introduction of memory, and the diffusive limit of the ISF is reached well within $20\,$ps. However, the trajectory simulated with memory friction is smoother than the Langevin trajectory. 
Therefore, on short timescales the GLE ($\omega_{c}=1.0\,$ps$^{-1}$) dynamics are closer to ideal ballistic motion, in comparison to the Langevin dynamics in which the particle undergoes more frequent changes of direction. The smoothing effect is captured mathematically by the ISF via the balance of decay amplitudes in the short-time, ballistic limit and the long-time, diffusive limit.

\section{Conclusions}

We have derived analytical intermediate scattering functions (ISFs) for a classical adsorbate obeying the Generalised Langevin Equation (GLE) in an unconfined one-dimensional landscape, with exponential memory friction. The expression is a product form, where each factor is a collapsing exponential of the form familiar from the well known memory-free (Langevin) result. In the unconfined case, a key feature of the ISF derived from the GLE is that the amplitude of the diffusive tail in the long time limit is reduced \textcolor{blue}{by a factor that depends exponentially on the decay time $1/\omega_{c}$ of the friction kernel}, while its decay rate is unchanged such that the diffusion coefficient is not affected. The variation of the diffusive tail amplitude with the cutoff frequency in the friction kernel can be interpreted in terms of the smoothing effect of memory friction on particle trajectories, i.e. realisations $x(t)$ of the GLE.

\section*{Acknowledgements}

The authors would like to thank Dr Bill Allison for helpful discussions. PT acknowledges the Engineering and Physical Sciences Research Council (EPSRC) for doctoral funding under the award reference 1363145.

\nolinenumbers

\section*{References}

\providecommand{\newblock}{}


\begin{thebibliography}{10}
	\expandafter\ifx\csname url\endcsname\relax
	\def\url#1{{\tt #1}}\fi
	\expandafter\ifx\csname urlprefix\endcsname\relax\def\urlprefix{URL }\fi
	\providecommand{\eprint}[2][]{\url{#2}}
	
	\bibitem{Cui2011ChemComm}
	Cui Y, Fu Q, Zhang H and Bao X 2011 {\em Chem. Commun.\/} {\bf 47}(5) 1470--72
	
	\bibitem{Hofmann2005PRL}
	Hofmann S, Cs{\'a}nyi G, Ferrari A~C, Payne M~C and Robertson J 2005 {\em Phys.
		Rev. Lett.\/} {\bf 95}(3) 036101
	
	\bibitem{Pawin2006Science}
	Pawin G, Wong K~L, Kwon K~Y and Bartels L 2006 {\em Science\/} {\bf 313} 961--2
	
	\bibitem{Kubo1966RepProgPhys}
	Kubo R 1966 {\em Rep. Prog. Phys.\/} {\bf 29}(1) 255--84
	
	\bibitem{MiretArtes2005JPCM}
	Miret-Art{\'{e}}s S and Pollak E 2005 {\em J. Phys-Condens. Mat.\/} {\bf 17}
	S4133--50
	
	\bibitem{Dai2010PRL}
	Dai J, Hou Y and Yuan J 2010 {\em Phys. Rev. Lett.\/} {\bf 104}(24) 245001
	
	\bibitem{Stanton2018PRX}
	Stanton L~G, Glosli J~N and Murillo M~S 2018 {\em Phys. Rev. X\/} {\bf 8}(2)
	021044
	
	\bibitem{Hanggi1990RevModPhys}
	H{\"{a}}nggi P, Talkner P and Borkovec M 1990 {\em Rev. Mod. Phys.\/} {\bf
		62}(2) 251--342
	
	\bibitem{Juaristi2008PRL}
	Juaristi I~J, Alducin M, Mui{\~{n}}o D~R, Busnengo F~H and Salin A 2008 {\em
		Phys. Rev. Lett.\/} {\bf 100}(11) 116102
	
	\bibitem{Rittmeyer2016PRL}
	Rittmeyer S~P, Ward D~J, G{\"u}tlein P, Ellis J, Allison W and Reuter K 2016
	{\em Phys. Rev. Lett.\/} {\bf 117}(19) 196001
	
	\bibitem{Gershinsky1996SurfSci}
	Gershinsky G, Georgievskii Y, Pollak E and Betz G 1996 {\em Surf. Sci.\/} {\bf
		365}(1) 159--67
	
	\bibitem{Rubin1963PhysRev}
	Rubin R~J 1963 {\em Phys. Rev.\/} {\bf 131}(3) 964--89
	
	\bibitem{Maekawa1980PLA}
	Maekawa M and Wada K 1980 {\em Phys. Lett. A\/} {\bf 80}(4) 293--5
	
	\bibitem{Mokshin2005NJP}
	Mokshin A~V, Yulmetyev R~M and H{\"{a}}nggi P 2005 {\em New Journal of
		Physics\/} {\bf 7}(9)
	
	\bibitem{MartinezCasado2007JPCM}
	Mart{\'{i}}nez-Casado R, Vega J~L, Sanz A~S and Miret-Art{\'{e}}s S 2007 {\em
		J. Phys. Condens. Mat.\/} {\bf 19}(30) 305002
	
	\bibitem{Harp1970PRA}
	Harp G~D and Berne B~J 1970 {\em Phys. Rev. A\/} {\bf 2}(3) 975--96
	
	\bibitem{Gertner1989JCP}
	Gertner B~J, Wilson K~R and Hynes J~T 1989 {\em J. Chem. Phys.\/} {\bf 90}
	3537--58
	
	\bibitem{Grote1980JCP}
	Grote R~F and Hynes J~T 1980 {\em J. Chem. Phys.\/} {\bf 73}(6) 2715--32
	
	\bibitem{Hershkovitz1999SurfSci}
	Hershkovitz E, Talkner P, Pollak E and Georgievskii Y 1999 {\em Surf. Sci.\/}
	{\bf 421}(1--2) 73--88
	
	\bibitem{Huang2011NatPhys}
	Huang R, Chavez I, Taute K~M, Luki{\'{c}} B, Jeney S, Raizen M~G and Florin E~L
	2011 {\em Nat. Phys.\/} {\bf 7} 576--80
	
	\bibitem{Pusey2011Science}
	Pusey P~N 2011 {\em Science\/} {\bf 332} 802--3
	
	\bibitem{Jardine2009ProgSurfSci}
	Jardine A~P, Hedgeland H, Alexandrowicz G, Allison W and Ellis J 2009 {\em
		Prog. Surf. Sci.\/} {\bf 84}(11--12) 323--79
	
	\bibitem{Lechner2015PCCP}
	Lechner B~A~J, Hedgeland H, Jardine A~P, Allison W and Ellis J 2015 {\em Phys.
		Chem. Chem. Phys.\/} {\bf 17}(34) 21819--23
	
	\bibitem{Hedgeland2009NatPhy}
	Hedgeland H, Fouquet P, Jardine A~P, Alexandrowicz G, Allison W and Ellis J
	2009 {\em Nat. Phys.\/} {\bf 5}(8) 561--4
	
	\bibitem{Vega2004JPCM}
	Vega J~L, Guantes R and Miret-Art{\'{e}}s S 2004 {\em J. Phys-Condens. Mat.\/}
	{\bf 16}(29) S2879--94
	
	\bibitem{Vega2004JCP}
	Vega J~L, Guantes R, Miret-Art{\'{e}}s S and Micha D~A {\em J. Chem. Phys.\/}
	{\bf 121}(17) 8580--88
	
	\bibitem{MartinezCasado2010ChemPhys}
	Mart{\'{i}}nez-Casado R, Sanz A~S, Vega J~L, Rojas-Lorenzo G and
	Miret-Art{\'{e}}s S 2010 {\em Chem. Phys.\/} {\bf 370}(1--3) 180--93
	
	\bibitem{Chudley1961PPSL}
	Chudley C~T and Elliott R~J 1961 {\em P. Phys. Soc. Lon.\/} {\bf 77}(494)
	353--61
	
	\bibitem{MartinezCasado2007JCP}
	Mart{\'{i}}nez-Casado R, Vega J~L, Sanz A~S and Miret-Art{\'{e}}s S 2007 {\em
		J. Chem. Phys.\/} {\bf 126}(19) 194711
	
	\bibitem{Grabert1988PhysRep}
	Grabert H, Schramm P and Ingold G~L 1988 {\em Physics Reports\/} {\bf 168} 115
	-- 207
	
	\bibitem{Berkowitz1983JCP}
	Berkowitz M, Morgan J~D and McCammon J~A 1983 {\em J. Chem. Phys.\/} {\bf
		78}(6) 3256--61
	
	\bibitem{Wan1998MolPhys}
	Wan S~Z, Wang C~X and Shi Y~Y 1998 {\em Mol. Phys.\/} {\bf 93}(6) 901--12
	
	\bibitem{Bao2004JStatPhys}
	Bao J~D 2004 {\em J. Stat. Phys.\/} {\bf 114}(1--2) 503--513
	
	\bibitem{Stella2014PRB}
	Stella L, Lorenz C~D and Kantorovich L 2014 {\em Phys. Rev. B\/} {\bf 89}(13)
	134303
	
	\bibitem{Ceriotti2011JCP}
	Ceriotti M, Manolopoulos D~E and Parrinello M 2011 {\em J. Chem. Phys.\/} {\bf
		134}(8) 084104
	
	\bibitem{Ianconescu2015JCP}
	Ianconescu R and Pollak E 2015 {\em J. Chem. Phys.\/} {\bf 143}(10) 104104
	
	\bibitem{Weiss2012QDS}
	Weiss U 2012 {\em Quantum Dissipative Systems\/} 4th ed ({\em Series in Modern
		Condensed Matter Physics\/} vol~13) (World Scientific) ISBN 10 981-4374-91-1
	
	\bibitem{Riley2006MathMeth}
	Riley K, Hobson M and Bence S 2006 {\em Mathematical Methods for Physics and
		Engineering\/} 3rd ed (Cambridge University Press) ISBN 978-0-521-67971-8
	
	\bibitem{Berne1966JCP}
	Berne B~J, Boon J~P and Rice S~A 1966 {\em J. Chem. Phys.\/} {\bf 45}(4)
	1086--96
	
	\bibitem{Caratti1998ChemPhysLett}
	Caratti G, Ferrando R, Spadacini R, Tommei G~E and Zelenskaya I 1998 {\em Chem.
		Phys. Lett.\/} {\bf 290}(4--6) 509--13
	
	\bibitem{Ottobre2011Nonlin}
	Ottobre M and Pavliotis G~A 2011 {\em Nonlinearity\/} {\bf 24}(5) 1629--53
	
	\bibitem{Nose1984JCP}
	Nos{\'e} S 1984 {\em J. Chem. Phys.\/} {\bf 81} 511--519
	
	\bibitem{Hoover1985PRL}
	Hoover W~G 1985 {\em Phys. Rev. A\/} {\bf 31}(3) 1695--1697
	
	\bibitem{Bussi2007JCP}
	Bussi G, Donadio D and Parrinello M 2007 {\em J. Chem. Phys.\/} {\bf 126}
	014101
	
	\bibitem{Basconi2013JCTC}
	Basconi J~E and Shirts M~R 2013 {\em J. Chem. Theor. Comp.\/} {\bf 9}
	2887--2899
	
\end{thebibliography}
\end{document}